\documentclass[journal=jctcce,manuscript=article]{achemso}

\usepackage{graphicx}
\usepackage{amsmath,amssymb}
\usepackage{xcolor}
\usepackage[version=4]{mhchem}
\usepackage{physics}
\usepackage{siunitx}
\AtBeginDocument{\RenewCommandCopy\qty\SI}

\newcommand{\bX}{{\bf X}}
\newcommand{\bY}{{\bf Y}}

\author{Jennifer R. DeRosa}
\email{jderosa@sas.upenn.edu}
\affiliation{Department of Chemistry, University of Pennsylvania, Philadelphia, Pennsylvania 19104, United States}
\alsoaffiliation{Department of Chemistry, Princeton University, Princeton, New Jersey 08544, United States}

\author{Tian Qiu}
\affiliation{Department of Chemistry, Princeton University, Princeton, New Jersey 08544, United States}
\alsoaffiliation{Department of Chemistry, University of Pennsylvania, Philadelphia, Pennsylvania 19104, United States}

\author{D. Vale Cofer-Shabica}
\affiliation{Department of Chemistry, Princeton University, Princeton, New Jersey 08544, United States}
\alsoaffiliation{Department of Chemistry, University of Pennsylvania, Philadelphia, Pennsylvania 19104, United States}

\author{Joseph E. Subotnik}
\alsoaffiliation{Department of Chemistry, University of Pennsylvania, Philadelphia, Pennsylvania 19104, United States}
\affiliation{Department of Chemistry, Princeton University, Princeton, New Jersey 08544, United States}

\title[]{Marcus Theory and The Condon Approximation Revisited II: The Horror of Triplet Energy Transfer}

\begin{document}

\begin{tocentry}

\begin{center}
\includegraphics[height=4cm]{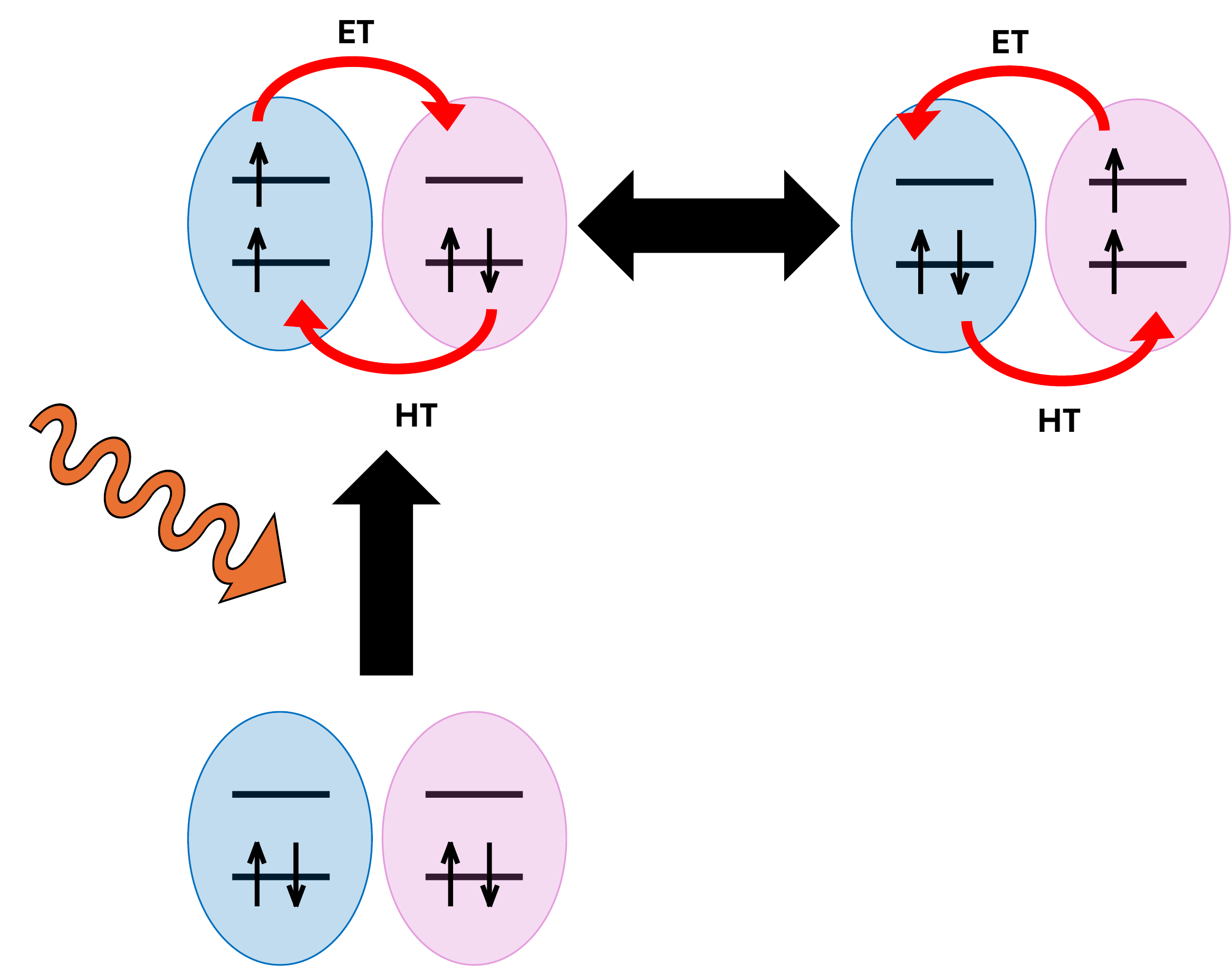}
\end{center}

\end{tocentry}

\begin{abstract}
We investigate the applicability of the Condon approximation (i.e. the notion that the diabatic coupling is invariant to geometry) in the context of both electron transfer (ET) and triplet energy transfer (TET) and compare the two cases.
Although it is well appreciated that diabatic couplings usually arise from the interactions of electronic wavefunction tails, 
we show that ET tails are very different from TET tails.  
Using a simple model problem, our analysis explains in detail why  the rates of TET decays with twice the rate of ET, while also leading to the hypothesis that the smaller diabatic couplings found for TET (versus ET) should imply more sensitivity to non-Condon fluctuations.  As an example, for the classic sets of molecules investigated by Closs, we show that the Condon approximation is indeed less applicable for TET than for ET.
\end{abstract}

\section{Introduction}

\subsection{Electron vs Triplet Energy Transfer}

\subsubsection{Rate Relation}

\begin{figure*}
    \centering
    \includegraphics[width=\textwidth]{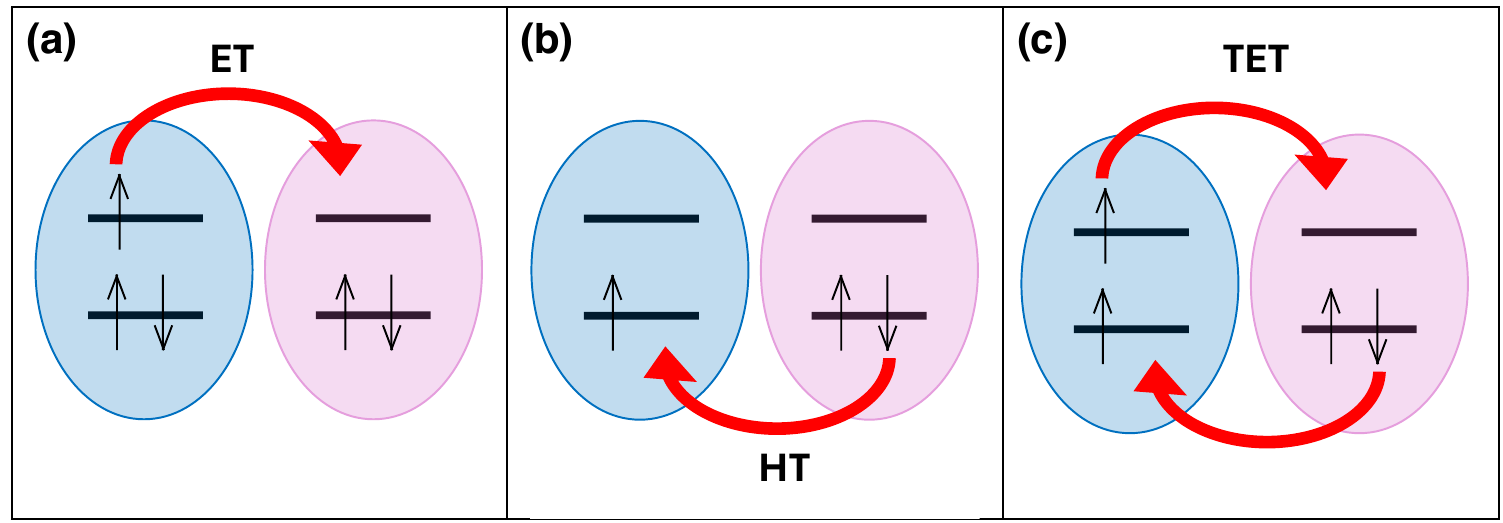}
    \caption{\label{fig:et_ht_eet_drawing}
    Schematic MO diagram of two molecules (donor, blue; acceptor, pink) containing two relevant orbitals (horizontal lines) with electrons (up/down arrows signifying spin) moving (thick red arrow) for electron transfer (ET), hole transfer (HT), and triplet energy transfer (TET) from donor to acceptor. 
    }
\end{figure*}

Rates of both electron transfer (ET) and triplet energy transfer (TET), $k_{\mathrm{ET}}$ and $k_{\mathrm{TET}}$, decay exponentially with the distance between donor and acceptor.  Within the ET literature, this decay is often quantified by the Greek letter $\beta$, and one writes $k_{\mathrm{ET}} = k_{\mathrm{ET}}^0 e^{-\beta_{\mathrm{ET}} R_{\textrm{DA}} }$;  similarly one can write, $k_{\mathrm{TET}} = k_{\mathrm{TET}}^0 e^{-\beta_{\mathrm{TET}} R_{\textrm{DA}} }$.   Historically, a great deal of energy and scientific interest has been focused on understanding the similarities between these two different nonadiabatic phenomena. On the one hand, ET requires only a single electron (1e) to transfer from donor (D) to an acceptor (A). On the other hand, TET is a two electron (2e) event where the triplet \emph{excitation} is transferred from D to A. Note, though, that TET can also be understood as the product of a single hole transfer (HT) and a single electron transfer (ET) (see Figure \ref{fig:et_ht_eet_drawing}).
As such, the argument can be made that, as a function of the donor-acceptor distance, the TET rate should scale as square of the ET rate, ie.     $\beta_{\mathrm{TET}}  = 2\beta_{\mathrm{ET}},$ 
$\beta_{\mathrm{TET}}  = 2\beta_{HT},$ or
\begin{eqnarray}
\label{eq:scale}
\beta_{\mathrm{TET}}  = \beta_{\mathrm{ET}} + \beta_{HT}.
\end{eqnarray}

\subsubsection{Measurements on Closs series}

\begin{figure}[htp!]
    \centering
    \includegraphics[width=\linewidth]{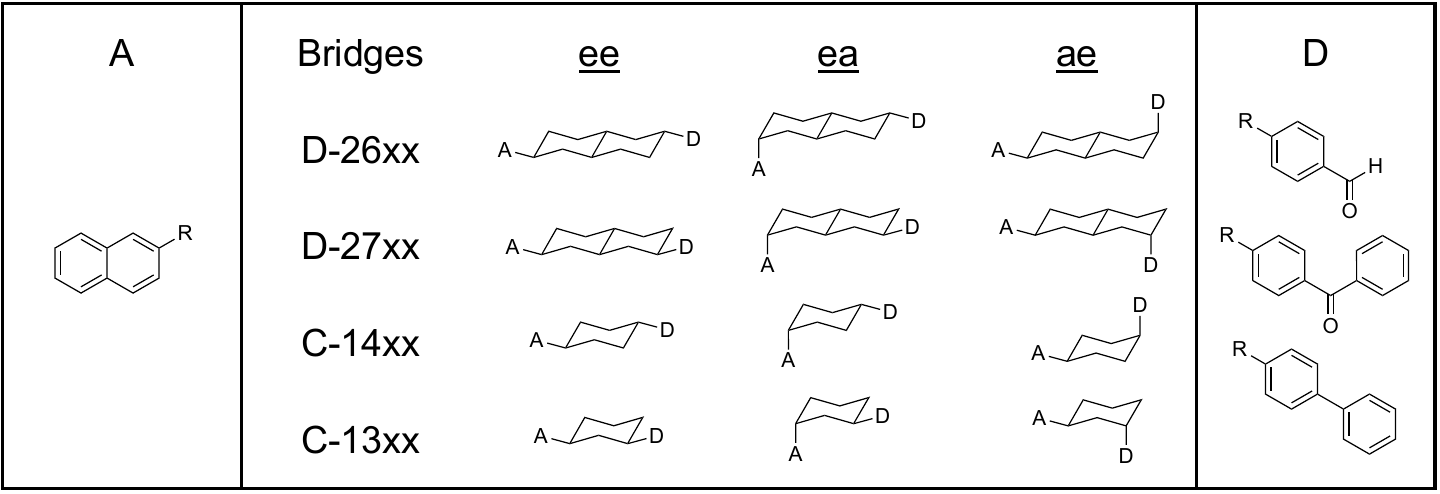}
    \caption{
    The Closs donor--bridge--acceptor molecules; 
    The donor (D) molecules from top to bottom: benzaldehyde (TET calculations), benzophenone (TET experiments), and biphenyl (ET/HT). The acceptor molecule (A) is naphthalene. The two bridge molecules: decalin (D-\#\#xx) and cyclohexane (C-\#\#xx). The D and A are bound at either the 1,3 or 1,4 position of the cyclohexane bridge (C-13xx/C-14xx) and the 2,6 and 2,7 position of the decalin bridge (D-26xx and D-27xx). The D and A are covalently bonded to the bridge either equatorially (e) or axially (a).
    }
\label{fig:molecule}
\end{figure}

\begin{figure}[htp!]
    \centering   \includegraphics[width=\linewidth]{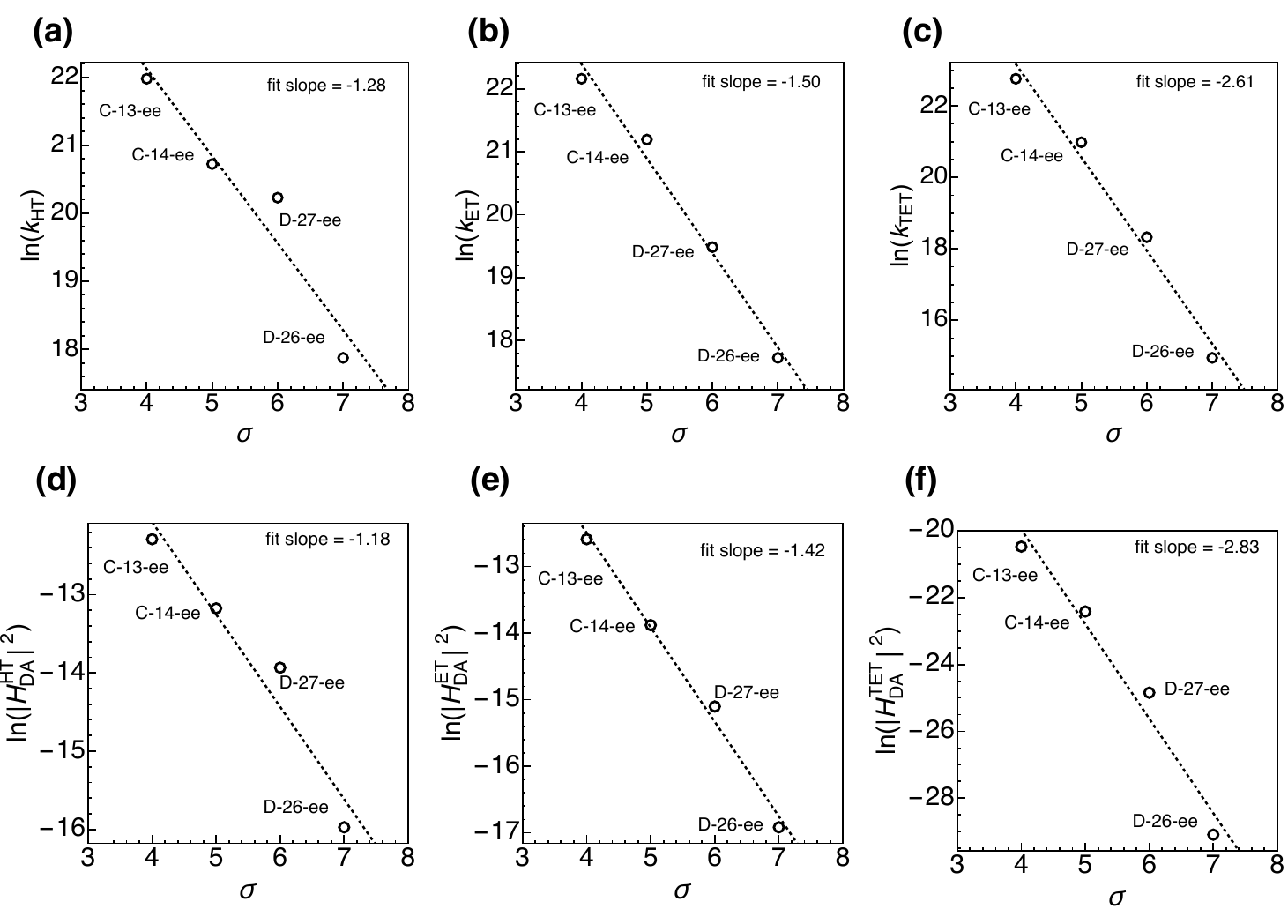}
    \caption{(a-c) Semi-logarithmic plots of experimental rates as measured from Ref. \citenum{closs1988intramolecular} and \citenum{closs1988EET} for the `ee' Closs molecules versus number of sigma bonds separating the D-A moieties. Note that the rate of triplet transfer falls off  exponentially with only small differences.
    (d-f) The computationally predicted squares of the diabatic coupling (Hartrees).  The data for TET in Fig. (d) is from Ref. \citenum{subotnik2010closs}. We calculate the ET (e) and HT (f) using (eDSC/hDSC) see Sec \ref{sec:discussion-calculations}.  
   All theoretical calculated used the optimized geometries from a Hartree-Fock singlet ground state calculation of the neutral species.    Note that theory matches experiment closely and also falls off exponentially.}
\label{fig:closs_exp}
\end{figure}

Several years ago now, the above claim was verified by Closs and co-workers while studying a series of donor-bridge-acceptor (D-B-A) systems.\cite{closs:1986:jcp:inverted,closs1989connection,closs1988EET,closs1988intramolecular} In their work, the donor (benzophenone/biphenyl) and acceptor (naphthalene) are covalently bonded to a series of rigid bridges as shown in Figure \ref{fig:molecule}.
For convenience, we plot the measured rates from the literature for the Closs molecules in Figure \ref{fig:closs_exp}a-c. Note that we only include the `ee' molecules, meaning both D and A are covalently bonded to the carbon of the bridge at the equatorial site. The semi-logarithmic plot for ET (b) and TET (c) show a linear dependence on the D-A distance according to the linear fit: adding the HT (-1.28) and ET (-1.50) slopes approximately equals the slope of the TET (-2.61) line.

\subsection{The Initial and Final States of ET and TET as Viewed Through the Marcus Process}
The experimental graphs in Figure \ref{fig:closs_exp} are quite convincing insofar as reproducing exponential decay of TET and confirming Eq. \ref{eq:scale}. Moreover,  from a theoretical perspective, if one performs calculations using the techniques described in Secs. \ref{sec:theory-and-results:1D}-\ref{sec:discussion} below, one again finds exponential decay and rates that satisfy Eq. \ref{eq:scale}.  Nevertheless, from a mathematical point of view, one must admit that  underlying the quadratic scaling is not entirely obvious--especially if one seeks an electronic structure framework that goes beyond single electrons (and includes many-body electron-electron correlation).

To appreciate the nuances of the scaling question, note that historically almost all discussions of ET have been based on Marcus theory which states that, 
in the nonadiabtic high-temperature limit, the rate of ET between an initial (D) and final (A) state is given by:\cite{nitzanbook}
\begin{equation}\label{eq:rate}
    k_{\textrm{D} \rightarrow \textrm{A}} = \frac{2 \pi}{\hbar} |H_{\textrm{DA}}|^2 \sqrt{\frac{1}{4 \pi k_B T \lambda}} e^{-\frac{(\lambda + \Delta G^0)^2}{4 \lambda k_B T}}
\end{equation}
$H_{\textrm{DA}}$ is the electronic diabatic coupling between the two states, and $\Delta G^0$ is the total free energy change between initial and final states. $\lambda$ is the reorganization energy. $\hbar$ and $k_B$ are the reduced Planck's constant and the Boltzmann constant, respectively. 

In the early days of ET calculations, before modern electronic structure theory was applied, one usually approximated the initial and final states as being one electron orbitals\cite{newton1991quantum}, say $\ket{d}$ and $\ket{a}$, so that 
the diabatic coupling ($H_{\textrm{DA}}$) was approximated as being proportional to the overlap of those two orbitals, $H^{\mathrm{ET}}_{\textrm{DA}} \propto S_{\textrm{da}}$.
Thereafter, given the intuition in Fig.~\ref{fig:et_ht_eet_drawing}, one can further suppose that, if TET is a 2e process with  two simultaneous 1e transfers of different spins, then $H^{\mathrm{TET}}_{\textrm{DA}} \propto S_{ad}^2$.

While the argument above appears quite reasonable at first glance, many questions inevitably arise when one attempts to go beyond the single electron picture and use modern electronic structure theory. After all, $H^{\mathrm{TET}}$ and $H^{\mathrm{ET}}$ have the same units (energy), so how can one matrix element be proportional to the square of the other? What is the relevant constant that emerges? Intuitively, if we think about individual electronic processes, one might presume that, in the spirit of superexchange\cite{nitzanbook}, one could  estimate:
\begin{equation}\label{eq:superexhange}
    H^{\mathrm{TET}}_{\textrm{DA}} = \frac{H^{\mathrm{ET}}_{\textrm{DA}}  H^{HT}_{\textrm{DA}}}{\Delta E},
\end{equation}
where $\Delta E$ is the gap between the ground state and a high lying charge-transfer virtual state. Indeed, we will revisit this successive single-electron expression for TET in Sec. \ref{sec:discussion} in the context of a multi-electronic wavefunction.   More generally, it is well-known that triplet energy states interact directly through two-electron dexter transfer terms (rather than successive one-electron perturbations). Mathematically,  if one assumes a ground state of the form $\ket{G} = \ket{d\bar{d}a\bar{a}}$, a donor diabatic state of the form $\ket{D} = \ket{d d^* a \bar{a}}$, and an acceptor diabatic state of the form $\ket{A} = \ket{d\bar{d}aa^*}$, then the diabatic coupling is nothing more than the exchange matrix element:
\begin{eqnarray}
    \bra{D}H\ket{A} =  \left<d^*\bar{a}\middle|\frac{1}{r_{12}}\middle|\bar{d}a^*\right> = 
-\left(d^*a^*\middle|\bar{d}\bar{a}\right)
\label{eq:naive}
\end{eqnarray}
Using modern electronic structure theory,
there is no easy means to relate Eq.~\ref{eq:naive}  to something like Eq.~\ref{eq:superexhange}, where the TET diabatic coupling is the product of electron transfer and hole transfer events.  

Herein lies the goal of this letter: to use modern excited state theory to explore in detail how exactly the scaling law above (e.g. Eq. \ref{eq:scale}) arises.
Before going any further, however, a few words are necessary regarding how to construct the initial and final electronic states for ET and TET. 

\subsubsection{Diabatization via Adiabat to Diabat Transformation}
It is important to emphasize that, if we want to include electron-electron correlation, we need to work with adiabats and then perform an adiabat to diabat (ATD) transformation.\cite{pacher1988approximately,pacher1993adiabatic}
While there is no unique, global solution for forming diabatic states,\cite{mead1982conditions}
in the case of interacting excited-state triplets, diabatic matrix elements between CIS triplets can be directly calculated using a variety of localized diabatization schemes\cite{subotnik:2015:acr}, including the fragment spin difference method\cite{voityuk2002fragment,hsu2008characterization,hsu:2010:fsd},  BoysOV diabatization\cite{subotnik2008boys,subotnik2009initial}, and others.  Given that BoysOV has already been successfully benchmarked against TET energy transfer for the Closs molecules\cite{subotnik2010closs}, we will work with Boys states\cite{subotnik2008boys,subotnik2009initial,foster1960canonical} for ET and HT,  and BoysOV states below for TET.  As a brief reminder, for any diabatization scheme with two adiabatic states ($\ket{\Phi_1}$ and $\ket{\Phi_2}$), one generates diabatic states by mixing the adiabats
\begin{equation}
\begin{pmatrix}\label{eq:diab-roation}
    \ket{\Xi_D}\\
    \ket{\Xi_A}\\
\end{pmatrix}
=
\begin{pmatrix}
        \phantom{-}\cos\theta & \sin\theta \\
        -\sin\theta & \cos\theta
\end{pmatrix}
\begin{pmatrix}
    \ket{\Phi_1}\\
    \ket{\Phi_2}\\
\end{pmatrix}
\end{equation}
where $\theta$ is the angle for the rotation matrix $U$ that forms TET states.
The final diabatic coupling is of the form:
\begin{equation}\label{eq:boys-coupling}
    H_{\textrm{DA}} = \frac{1}{2} \sin2\theta |\epsilon_{2} - \epsilon_{1}|
\end{equation}
The Boys recipe for generating diabatic states is based on an imaginary solvent that exerts a reasonably strong electric field that mixes adiabats together and favors charge localization;\cite{subotnik2009initial} this method is appropriate for charge transfer and is achieved by maximizing Eq. \ref{eq_fboys}.
\begin{align}\label{eq_fboys}
    f_{\textrm{Boys}}({\Xi_i}) = \sum_{i,j \in \{\textrm{A, D}\}} |\bra{\Xi_i} &\vec{\mu} \ket{\Xi_i} - \bra{\Xi_j} \vec{\mu} \ket{\Xi_j} |^2 
\end{align}
For TET, one must slightly alter the argument as there is no charge dipole, and while one could certainly use Edmiston-Ruedenberg localization\cite{rued:1963,subotnik2009initial}, a cheaper alternative is to use occupied-virtual boys (or BoysOV) localization which seeks to localize the particle and hole for an excited state calculation; in other words, one maximizes the function:
\begin{align}\label{eq:boys-function}
    f_{\textrm{BoysOV}}({\Xi_i}) = \sum_{i,j \in \{\textrm{A, D}\}} |\bra{\Xi_i} &\vec{\mu}^{occ} \ket{\Xi_i} - \bra{\Xi_j} \vec{\mu}^{occ} \ket{\Xi_j} |^2 + \nonumber
    \\&|\bra{\Xi_i} \vec{\mu}^{vir} \ket{\Xi_i} - \bra{\Xi_j} \vec{\mu}^{vir} \ket{\Xi_j} |^2
\end{align}
Here, $\vec{\mu}^{occ}$ and $\vec{\mu}^{virt}$ are the occupied-occupied and virtual-virtual components of the dipole tensor. Note that 
Eq.~\ref{eq:boys-function} is most naturally applicable for CIS or TD-DFT/TDA states as they can be defined in terms of virtual particles and occupied holes\cite{mhg1995attachdetach}.

\subsubsection{The Ubiquitous Electronic Structure Problem}
The problem of how to generate the original adiabats still remains. While TET can certainly be calculated with a combination of Hartree-Fock and CIS theory\cite{subotnik:2015:acr}, ET for a radical species is far more nuanced. After all, this process occurs primarily between the ground and excited state, which precludes the use of simple `black box' methods like HF/CIS to model the electronic structure of the system.
Brillouin's theorem states that singly excited determinants will not interact directly with a reference HF determinant.\cite{szabo:ostlund}
Therefore, one must go beyond HF/CIS for the ground and excited state to be `on the same footing' such that a diabatic coupling can be computed and used in Eq.~\ref{eq:rate}.
Multireference methods such as Complete Active Space Self-Consistent Field (CASSCF)\cite{roos1980complete,siegbahn1981complete}, Multi-Reference Configuration Interaction (MRCI)\cite{martinez1996multi,grimme1999combination}, and Multi-Reference Perturbation Theory (MRPT)\cite{hirao1992multireference} resolve the aforementioned issue, but they come with their own set of challenges; namely, the solution is highly dependent on how one selects the active space, so it is not guaranteed that the solutions will capture the two states for ET.
Previous work in our research group developed a dynamically weighted (DW), state-averaged (SA), constrained CASSCF algorithm (eDSC/hDSC) to specifically handle ET/HT for radical systems,\cite{qiu2024fast} and will be invoked below.

\subsection{Path Forward and Outline}
To reiterate the goal of this work, we would like to understand in detail (and quantitatively) the relationship between ET and TET, and the origin of the quadratic scaling in Eq.~\ref{eq:rate} above. In this regard, we will build a simple one-dimensional (1D) model which allows for both ET and TET dynamics, and we will compare the rates for both. We will show that, as one might have guessed, the scaling law comes from the existence of tails that are inevitably present when we build diabatic states from an ATD transformation. In other words, to zeroth order and in a very loose sense, let 
$\ket{a},\ket{b}$ be one electron orbitals entirely localized to the acceptor; 
let 
$\ket{d},\ket{e}$ be one electron orbitals entirely localized to the donor;
let 
$\ket{\Phi_0}$ represent a wavefunction for a closed-shell collection of stationary (inactive) electrons. 
Then, for ET, we can often approximate that the relevant many-body diabatic states are of the form 
\begin{eqnarray}
\ket{D_{\mathrm{ET}}} = \ket{\Phi_0 d a \bar{a}} \\
\ket{A_{\mathrm{ET}}} = \ket{\Phi_0 d \bar{d} a}  
\end{eqnarray}
when looking to represent the initial and final states for ET.
For TET, however, we will find that the relevant diabatic states necessarily have a more complicated form, e.g. an $m_s = +1$ state of the form:
\begin{eqnarray}\label{eq:DTET_tails_guess}
\ket{D_{\mathrm{TET}}} = \alpha \ket{\Phi'_0 dea\bar{a}} +
\beta \ket{\Phi'_0 db a \bar{a}} 
+ \gamma \ket{\Phi'_0 d \bar{d} e a }
+ \delta \ket{\Phi'_0 d \bar{d} ab} 
\end{eqnarray}
with $\alpha \gg \beta,\gamma,\delta$.  
The tails\cite{vura2010characterizing} proportional to $\beta,\gamma,\delta$ are the critical terms as far as establishing the final diabatic coupling (that decays with $\beta_{\mathrm{TET}} = 2\beta_{\mathrm{ET}}$), and some tails are more important than others (as seen below).  The fact that such tails are so small and inevitably sensitive to environment would seem to imply that TET violates the Condon approximation more regularly than does ET or HT.

As an interesting application of the finding above, we revisit the Closs molecules\cite{closs1988intramolecular}, which were the original molecules demonstrating the scaling law in question between electron transfer (ET) and triplet excitation energy transfer (TET).\cite{closs1988EET,closs1989connection}
In a companion paper,\cite{eshake} we show that one data point (for the C-13ae molecule) was neglected in the original paper which falls far off the predicted $k=k_0e^{-\beta_{\mathrm{TET}} R}$ line. In fact, for the C-13ae molecule, Paper I shows that, as far as TET is concerned,  a  conical intersection (CI) is present.  Here, we will show that the presence of such a CI automatically ruins any scaling law between ET and TET, and we will hypothesize that, because TET involves two small tail interactions (rather than one for ET), a scaling law between ET and TET may be more rare than we think--given the strong possibility for non-Condon effects in TET.

\section{Theory and Results: 1-Dimensional Model}\label{sec:theory-and-results:1D}

\begin{figure}[htp!]
    \centering
    \includegraphics[width=\linewidth]{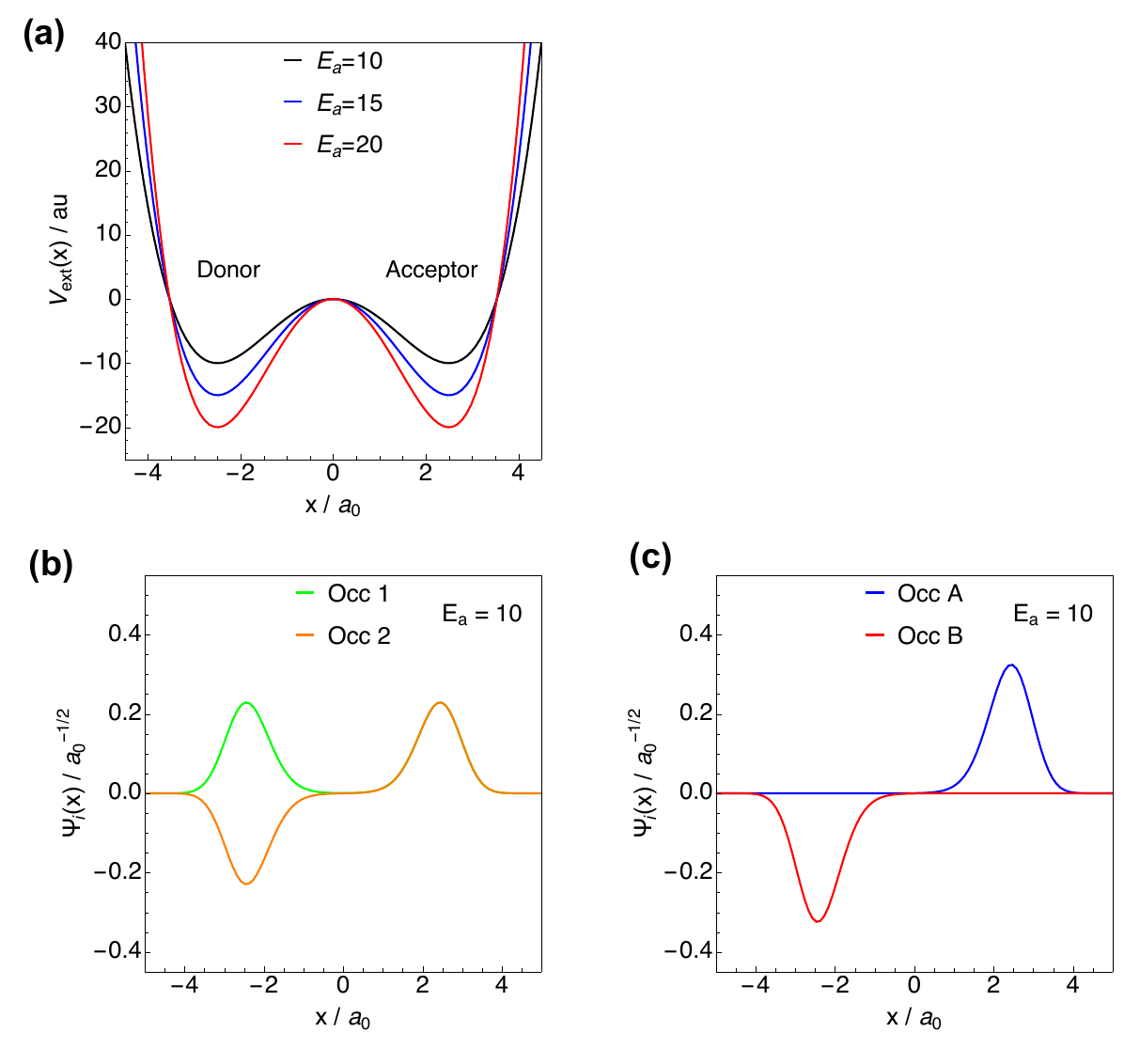}
    \caption{1D model of a 4-electron system using (a) $V_{ext}$ external potential as an approximate nuclear-electron attraction. The e-e repulsion term in the Hamiltonian is $ c_{ee} / \sqrt{|x_{12}|^2+a_h^2} $ where $a_h = 0.01 \, a_0$ and $ c_{ee} = \num{2.5e-5} $. (b) Occupied canonical (green; orange) and (c) localized (blue; red) orbitals from HF.
    }
    \label{fig:1d_toy_model_pot_orbs_etc}
\end{figure}

To probe the questions above and  quantify a putative tail effect, we have developed a straightforward 1D model systems that, unlike a more complicated 3D model or \emph{ab initio} systems, allows us to probe every single component of a diabatic state that contributes to the diabatic coupling.
We consider a 4-electron problem in 1-dimension on a grid as follows (atomic units assumed throughout):
 \begin{equation}\label{eq:H4e}
    H_{4e} = 
    - \sum_{i=1}^{4} \frac{1}{2} \frac{\partial^2}{\partial x_i^2}
    + V_{ext}
    + \sum_{i=1}^{4} \sum_{j>i}^{4} \frac{c_{ee}}{\sqrt{|x_{ij}|^2+a_h^2}}
 \end{equation}
\begin{equation}\label{eq:Vext}
    V_{ext} = \sum_{i=1}^{4}-8 \frac{ E_a}{L^2}x_i^2+16 \frac{E_a}{L^4}x_i
\end{equation}
Here, in order to mimic the potential energy for an electron that is attracted to both donor and acceptor atoms, the electron-nuclear coulomb attraction is modeled as an effective double-well potential $V_{ext}$ as shown in Figure \ref{fig:1d_toy_model_pot_orbs_etc}a. Below, we will probe varying barrier heights ($E_a$) of grid size $2L$. The ``donor'' well is the LHS ($x<0$) and the ``acceptor'' well on the RHS ($x>0$). Note that this model is symmetric, so the labeling of ``donor'' and ``acceptor'' is arbitrary. The electron-electron repulsion is modeled by the last term in Eq.~\ref{eq:H4e}, where we apply a shift to the distance between electrons i and j so that that the energy does not diverge -- unlike the standard Coulomb attraction that occurs in 3D. It is well known that one must avoid such discontinuities in 1D. We chose parameters for the electron-electron repulsion $a_h = 0.01 a_0$ and the dimensionless  constant $ c_{ee} = \num{2.5e-5} $.
 The problem is solved on a grid of $N_{\textrm{grid}}$ delta functions evenly spaced in position space from -5 to 5 $a_0$. We evaluate the diabatic coupling for a series of 30 different heights in Eq. \ref{eq:Vext} from 10 to 20 $E_h$.

\subsection{Electronic Structure}

\subsubsection{TET and HF/CIS}

In order to model TET, we followed canonical quantum chemistry and performed Hartree-Fock (HF)/configuration interactions singles (CIS) calculations. For the restricted Hartree-Fock (HF) equations, we work in our basis of delta functions and use a second-derivative five-point, fourth-order central finite difference stencil for the momentum operator. The initial guess uses the density matrix, $P$, from the converged SCF procedure of the next largest barrier height, and the largest barrier initial guess simply assumes two doubly-occupied orbitals from the 1e part of the Hamiltonian.
Once converged, we have in hand a set of orbitals: two doubly-occupied orbitals (Figure \ref{fig:1d_toy_model_pot_orbs_etc}b) and $N_{\textrm{grid}}-2$ virtual orbitals. The HF ground state wavefunction is:
\begin{equation}\label{eq:psi_hf}
    \ket{\Psi_{\mathrm{HF}}} = \ket{\Psi} = \ket{1 \bar{1} 2 \bar{2}}
\end{equation}
For CIS, we form a basis of single excitations from the ground-state Slater determinant such that an excited state is represented as a linear combination of that set ${\ket{\Psi_\mathbf{i}^\mathbf{a}}}$. 
\begin{equation}\label{eq:psi_cis}
    \ket{\Psi_{CIS}} = \sum_{\mathbf{i}}^{4} \sum_{\mathbf{a}}^{2N_{\textrm{grid}}-4} t_{\mathbf{i}}^{\mathbf{a}} \ket{\Psi_{\mathbf{i}}^{\mathbf{a}}}
\end{equation}
Where we use bold to indicate that $\mathbf{i}$ and $\mathbf{a}$ are occupied and virtual spin orbitals.
Here, $t_{\mathbf{i}}^{\mathbf{a}}$ are the  coefficients for the singly-excited Slater determinants $\ket{\Psi_\mathbf{i}^\mathbf{a}}$. 
We solve for the CIS coefficients by diagonalizing the CIS Hamiltonian.
\begin{equation}\label{eq:H_CIS}
    A_{\mathbf{iajb}} = F_{\mathbf{ab}}\delta_{\mathbf{ij}} - F_{\mathbf{ji}}\delta_{\mathbf{ab}} + \Pi_{\mathbf{ajib}} + E_{\mathrm{HF}}\delta_{\mathbf{ab}}\delta_{\mathbf{ij}}
\end{equation}
\begin{align}
    F_{\mathbf{pq}} = h_{\mathbf{pq}} + \sum_\mathbf{m} \Pi_{\mathbf{pmqm}}
\end{align}
Above, we have written the $A$ matrix in terms of the Fock matrix ($F_{pq}$ and 2e part, $\Pi$) and the HF ground state energy ($E_{\mathrm{HF}}$). Matrix $A_{\mathbf{iajb}}$ is $2N_{\mathrm{grid}}-4$ by $2N_{\mathrm{grid}}-4$ to account for both spin configurations $\alpha$ and $\beta$. Solving the eigensystem of $A_{\mathbf{iajb}}$ yields the CIS states and energies.
\begin{equation}
    \sum_{bj} A_{\mathbf{iajb}} X_{\mathbf{bj}} = E X_{\mathbf{ai}}
\end{equation}
Where the solutions are normalized according to $\sum_{\mathbf{ia}}X_{\mathbf{ai}}^*X_{\mathbf{ai}}=1$. $X_{\mathbf{ai}}$ specifies four spin diabats: a singlet and three triplets according to Eqs. \ref{eq:s0} through \ref{eq:t-1} below.
\begin{align}
    &X_{a_{\alpha}i_{\alpha}} = X_{a_{\beta}i_{\beta}} = \frac{1}{\sqrt{2}} si^a \label{eq:s0} \\
    &X_{a_{\alpha}i_{\alpha}} = -X_{a_{\beta}i_{\beta}} = \frac{1}{\sqrt{2}} t_i^{a(0)} \label{eq:t0} \\
    &X_{a_{\alpha}i_{\beta}} = t_i^{a(+1)} \label{eq:t1} \\
    &X_{a_{\beta}i_{\alpha}} = t_i^{a(-1)} \label{eq:t-1}
\end{align}
Note that $t_i^{a(0)} = t_i^{a(+1)} = t_i^{a(-1)}$. The two lowest energy solutions in a single triplet block, e.g. $t_i^{a(+1)}$, define the lowest two triplet adiabatic states, $\ket{\Phi_1}$ and $\ket{\Phi_2}$. 
From here, BoysOV diabatization (Eq.~\ref{eq:boys-function}) yields the D and A states 
$\ket{\Phi_D}$ and $\ket{\Phi_A}$.
as well as their corresponding coupling (Eq.~\ref{eq:boys-coupling}).

\subsubsection{HT and hDSC}\label{sec:1D-hDSC}
Next, we address our approach to ET and HT. We invoke the algorithm from Ref. \citenum{qiu2024fast}. In short, the algorithm begins by writing down a restricted open-shell wavefunction with a single unpaired electron for a $2n+1$ electron system.
\begin{equation}
    \ket{\Psi_1} = \ket{1\bar{1}, 2\bar{2}, ...,n\bar{n}, n+1}
\end{equation}
From here, one can either choose to model ET by exciting the unpaired electron to a higher energy orbital or to model HT by exciting the hole to a lower energy orbital, yielding the following excited state wavefunctions: ($\ket{\Psi_2^e}$ for ET and $\ket{\Psi_2^h}$ for HT)
\begin{equation}\label{eq:psiET}
    \ket{\Psi_2^e} = \ket{1\bar{1}, 2\bar{2}, ...,n\bar{n}, n+2}
\end{equation}
\begin{equation}\label{eq:psiHT}
    \ket{\Psi_2^h} = \ket{1\bar{1}, 2\bar{2}, ...,n, n+1 \overline{n+1} }
\end{equation}
In either case, one aims to minimize the weighted energy of the ground and excited state:
\begin{equation}\label{eq:Etot_weighted}
    E_{\rm tot}^{e/h} = w_1 \bra{\Psi_1} H \ket{\Psi_1} + w_2 \bra{\Psi_2^{e/h}} H \ket{\Psi_2^{e/h}},
\end{equation}
where $w_1$ and $w_2$ are dynamical weighting coefficients to balance the ground and excited state energies (see Ref. \citenum{qiu2024fast} for details). To avoid local excitation, an additional constraint is applied such that the vector space spanned by two active orbitals ($\phi_{n+1}$ and $\phi_{n+2}$ for ET, and $\phi_{n}$ and $\phi_{n+1}$ for HT) projects equally on to the vector space spanned by basis orbitals from two molecular fragments, e.g., the donor and acceptor. With such a constraint, the resulting two states must be a linear combination of a donor and acceptor diabatic state describing ET/HT.

In our simulations below, we apply hDSC to the system with 3 electrons, so that the two configurations read as follows:
\begin{align}
    \ket{\Psi^1_{\rm hDSC}} &= \ket{1\bar{1},2},\\
    \ket{\Psi^2_{\rm hDSC}} &= \ket{1,2\bar{2}}.
\end{align}
The constraint is defined such that the vector space of donor is the union of all delta functions centered at $x<0$, and the vector space of acceptor is the union of all delta functions centered at $x>0$. One can then build a Lagrangian to represent this constrained optimization problem:
\begin{align}
    \mathcal{L}_{\rm hDSC} = w_1\bra{\Psi^1_{\rm hDSC}}&\hat{H}\ket{\Psi^1_{\rm hDSC}}+w_2\bra{\Psi^2_{\rm hDSC}}\hat{H}\ket{\Psi^2_{\rm hDSC}} \\
    &-\lambda{\rm Tr}\left((\hat{P}_D-\hat{P}_A)(\ket{\phi_1}\bra{\phi_1}+\ket{\phi_2}\bra{\phi_2})\right),
\end{align}
where $\lambda$ is the Lagrange multiplier, $\hat{P}_D$ is the vector space spanned by the donor, and $\hat{P}_A$ is the vector space spanned by the acceptor. We solve the constrained optimization problem through an efficient DIIS-SQP algorithm\cite{qiu2024fast} to obtain two adiabatic states that represent the HT process. A Boys diabatization scheme (following Eq.~\ref{eq_fboys}) is then applied to find the corresponding diabatic states. Note that in the one-dimension case, such a transformation is equivalent to diagonalizing the $2\times2$ dipole matrix:
\begin{align}
    \begin{pmatrix}
        \bra{\Psi^1_{\rm hDSC}}x\ket{\Psi^1_{\rm hDSC}} &\bra{\Psi^1_{\rm hDSC}}x\ket{\Psi^2_{\rm hDSC}}\\
        \bra{\Psi^2_{\rm hDSC}}x\ket{\Psi^1_{\rm hDSC}} &\bra{\Psi^2_{\rm hDSC}}x\ket{\Psi^2_{\rm hDSC}}
    \end{pmatrix}.
\end{align} 
For more details, see Ref. \citenum{qiu2024fast}.

\subsection{Scaling Results}
\begin{figure}[htp!]
    \centering
    \includegraphics[width=\linewidth]{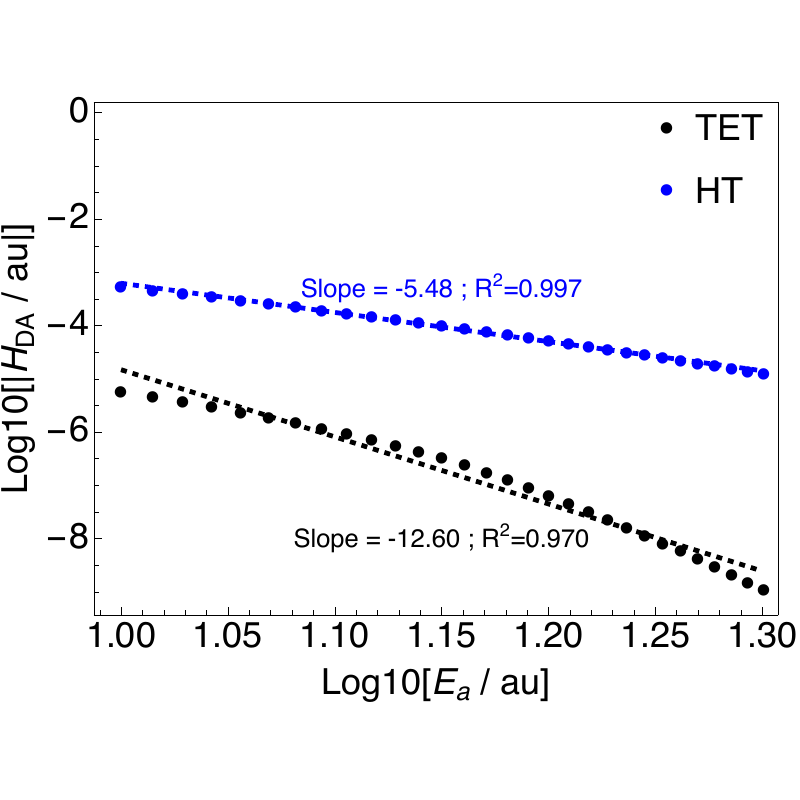}
    \caption{
    A log-log plot of the magnitude of the EET coupling of the 4-electron system from CIS boysOV diabats (black circles) and the HT coupling of the 3-electron system from hDSC (blue circles) with linear fit plotted in dashed lines. Note that the HT slope of -5.48 is approximately half that of the TET curve (-12.60).
    }
    \label{fig:loglog1D}
\end{figure}

Figure \ref{fig:loglog1D} shows a log-log plot of the diabatic coupling as a function of barrier heights, which range between 10 and 20 Hartree. For the parameter regimes of $a_h$ and $c_{ee}$ that we have tested, we observe the desired scaling or power relation between the TET coupling and HT  (a factor of ~2 difference in slopes for the dashed line fits: -12.60 TET/ -5.48 HT). The linear fit is better for the HT data (R\textsuperscript{2} = 0.997) than the TET (R\textsuperscript{2} = 0.970), which is slightly concave downwards. One might wonder whether the better linear fit for HT vs TET suggests that the nature of the TET  coupling is more sensitive to electronic structure than is the nature of the HT coupling?  Indeed,  our results below would suggest as much (and this will be discussed below). Nevertheless, one must be careful in making such an assertion because the hDSC algorithm for HT accounts for little correlation, while CIS theory does recover some.
In any event, after forming our two localized diabatic states for our 1D model, we can now perform further wavefunction analysis on the TET diabats to determine what parts of the wavefunction contribute to the diabatic coupling.

\subsection{Wavefunction Decomposition}
\begin{figure}[htp!]
    \centering
    \includegraphics[width=\linewidth]{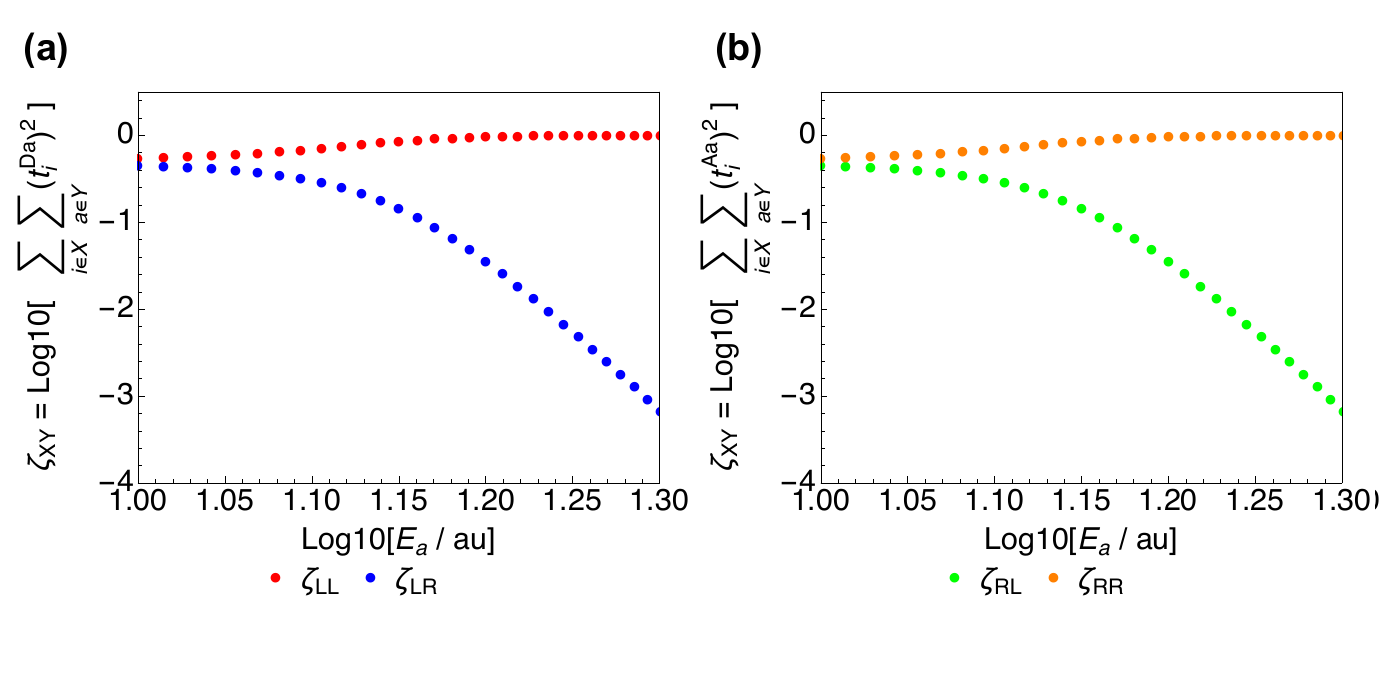}
    \caption{
    A log-log plot of the relatively important excitation decompositions  of (a) the donor wavefunction (L) and (b) acceptor wavefunction (R) as a function of the barrier height.
    }
    \label{fig:wavefunction-decomp}
\end{figure}

In order to glean as much insight as possible into the nature of the diabatic states, 
we begin by performing a change of basis within the occupied and virtual orbitals separately. Namely,  we  apply a unitary transformation from maximizing the BoysOV Eq. \ref{eq:boys-function}. In one dimension, this is the same as block diagonalizing the occupied and virtual orbital dipole matrix directly. We are left with orbitals that are primarily localized on either the left-hand, i.e. the donor ($x<0$) side or the right-hand, i.e. the acceptor ($x>0$) side of our 1D ``molecule'' (see Figure \ref{fig:1d_toy_model_pot_orbs_etc}c). We can then write our diabatic states in the basis of localized orbitals as  follows:
\begin{align}
    &\ket{\Psi_D} \equiv \ket{D} = \sum_{\tilde{i}}^{2} \sum_{\tilde{a}}^{N_{\textrm{grid}}-2} t_{\tilde{i}}^{D\tilde{a}} \ket{ \Psi_{\tilde{i}}^{\tilde{a}}} \\
    &\ket{\Psi_A} \equiv \ket{A} = \sum_{\tilde{i}}^{2} \sum_{\tilde{a}}^{N_{\textrm{grid}}-2} t_{\tilde{i}}^{A\tilde{a}} \ket{ \Psi_{\tilde{i}}^{\tilde{a}}}
\end{align}
Here $t_{\tilde{i}}^{D\tilde{a}}$ specifies the coefficients for an excitation of the donor diabat and $t_{\tilde{i}}^{A \tilde{a}}$ specifies the coefficients for an excitation of the the acceptor  diabats \emph{ in terms of the localized orbitals}, $\tilde{i}$ and $\tilde{a}$. There are four possible combinations of excitations that comprise the overall diabatic state.
\begin{align}\label{eq:wavefunction-decomp}
    \ket{\Psi_D} \equiv \ket{D} = 
    &  \sum_{\tilde{i} \in L}^{2} \sum_{\tilde{a} \in L}^{N_{\textrm{grid}}-2} t_{\tilde{i} }^{D\tilde{a}} \ket{ \Psi_{\tilde{i}}^{\tilde{a}}} 
    + \sum_{\tilde{i} \in L}^{2} \sum_{\tilde{a} \in R}^{N_{\textrm{grid}}-2} t_{\tilde{i} }^{D\tilde{a}} \ket{ \Psi_{\tilde{i}}^{\tilde{a}}} \\
    &+ \sum_{\tilde{i} \in R}^{2} \sum_{\tilde{a} \in L}^{N_{\textrm{grid}}-2} t_{\tilde{i} }^{D\tilde{a}} \ket{ \Psi_{\tilde{i}}^{\tilde{a}}} 
    + \sum_{\tilde{i} \in R}^{2} \sum_{\tilde{a} \in R}^{N_{\textrm{grid}}-2} t_{\tilde{i} }^{D\tilde{a}} \ket{ \Psi_{\tilde{i}}^{\tilde{a}}}
\end{align}
For simplicity, we introduce the following notation
\begin{equation}
    \ket{D_{L}^{R}} = \sum_{\tilde{i} \in L}^{2} \sum_{\tilde{a} \in R}^{N_{\textrm{grid}}-2} t_{\tilde{i} }^{D\tilde{a}} \ket{ \Psi_{\tilde{i}}^{\tilde{a}}}
\end{equation}
so that the cumbersome donor and acceptor diabat decomposition given in Eq.~\ref{eq:wavefunction-decomp} can be written succinctly.
\begin{equation}\label{eq:D_decomp}
    \ket{D} = \ket{D_L^L} + \ket{D_L^R} + \ket{D_R^L} + \ket{D_R^R}
\end{equation}
\begin{equation}\label{eq:A_decomp}
    \ket{A} = \ket{A_L^L} + \ket{A_L^R} + \ket{A_R^L} + \ket{A_R^R}
\end{equation}
Figure \ref{fig:wavefunction-decomp} plots the decomposition for the donor (Fig.~\ref{fig:wavefunction-decomp}a) and acceptor (Fig. \ref{fig:wavefunction-decomp}b). 
Unsurprisingly, $\ket{A}$ is primarily comprised of $\ket{A_R^R}$ excitations. $\ket{A_R^L}$ is comparable in contribution to the $\ket{A_R^R}$ for a barrier of 10, but decreases as the barrier is increased; $\ket{A_L^L}$ and $\ket{A_L^R}$ make up less than 0.001\% of the total wavefunction at any given barrier height and are excluded from Fig.~\ref{fig:wavefunction-decomp}.  Obviously, the same complimentary conclusions hold for $\ket{D}$. 

\begin{figure}[htp!]
    \centering
    \includegraphics[width=0.75\linewidth]{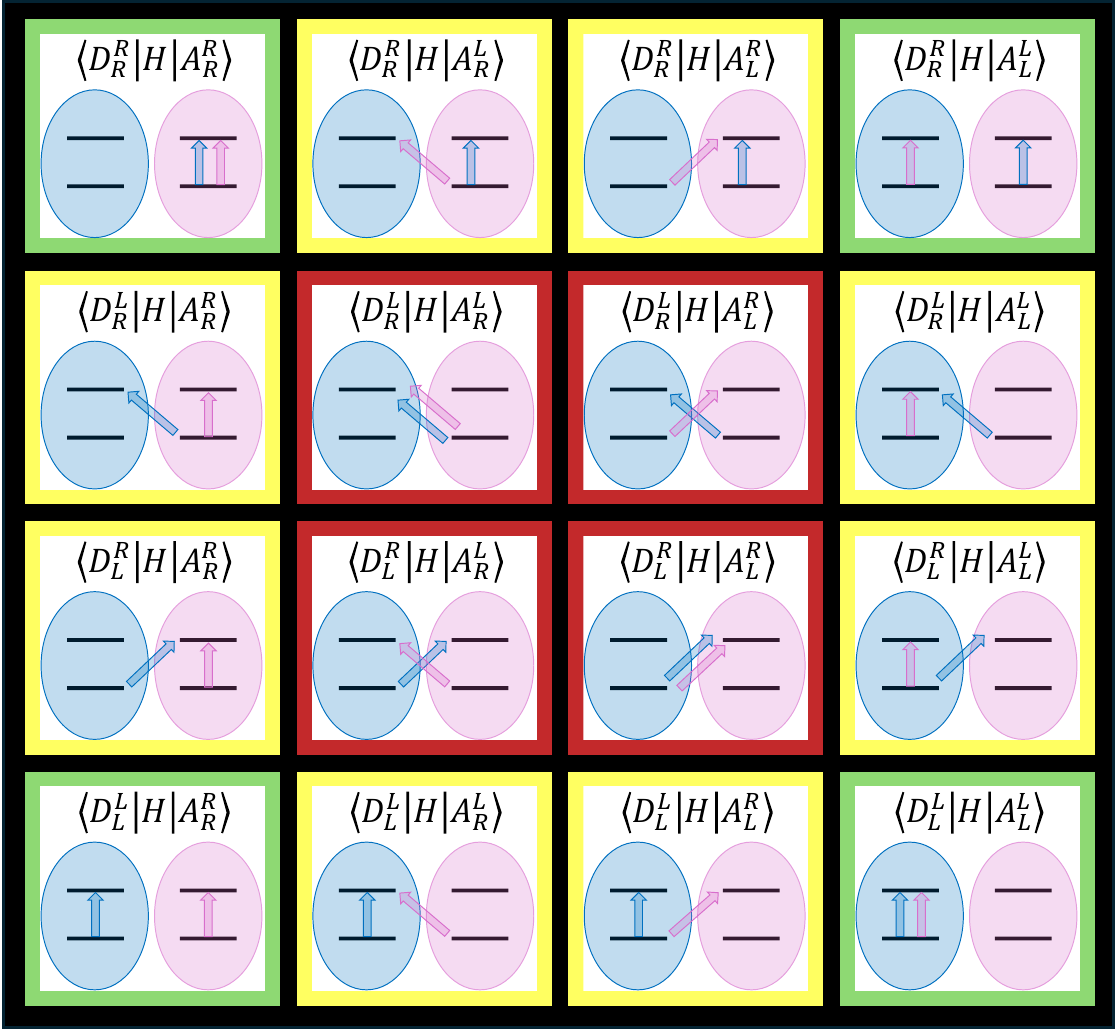}
    \caption{A table depicting schematics of the 16 terms that arise from decomposing the diabatic CIS wavefunctions $\bra{D}$ (blue) and $\ket{A}$ (pink). 
    These diabatic wavefunctions can have tails, and the diabatic coupling that arises between  $\bra{D}$ and $\ket{A}$ can involve exclusively local excitations (green), exclusively charge transfer excitations (red), or a combination of local and a charge transfer excitations (yellow) of the diabatic wavefunctions. Since $V_{ext}$ is a symmetric double well, the table is persymmetric such that term in position (i,j) is equal to the term in position (5-j,5-i). The sum of all 16 terms equals the total coupling.}
\label{fig:DXXHAXX_table}
\end{figure}

Next, we directly analyze the contribution of the tails to the diabatic coupling by evaluating $\bra{D}H\ket{A}$ with our decomposed wavefunctions for the donor and acceptor. From the decompostion of the CIS diabatic states from Eqs \ref{eq:D_decomp} and \ref{eq:A_decomp}, there are  16 terms in the diabatic coupling given by the sum
\begin{equation}\label{eq:diabH}
    \bra{D}H\ket{A} = \sum_{s,t \in \{R,L\}} \sum_{u,v \in \{R,L\}} \bra{D_{s}^{t}} H \ket{A_{u}^{v}}
\end{equation}
which is visualized in the 4x4 `matrix' shown in Figure \ref{fig:DXXHAXX_table}. As in Figure \ref{fig:et_ht_eet_drawing}, the blue oval depicts diabat $\ket{D}$ and pink the $\ket{A}$. Each element in the table visualizes one class of excitations. For example, row 4 column 1 visualizes the term $\bra{D_{L}^{L}} H \ket{A_{R}^{R}}$, where we include excitations from an occupied orbital (bottom rungs) localized on the left to a virtual orbital (top rungs) on the left in $\bra{D}$ and excitations from an occupied orbital localized on the right to a virtual orbital on the right in $\ket{A}$. Each excitation type for a single diabat can either be a local (vertical arrow) or charge transfer (diagonal) excitation, resulting in terms where exclusively local (green border) or exclusively charge transfer (red border) excitations of $\ket{D}$ and $\ket{A}$ are coupled together, as well as the cross terms (yellow). 

\begin{figure}[htp!]
    \centering
    \includegraphics[width=\linewidth]{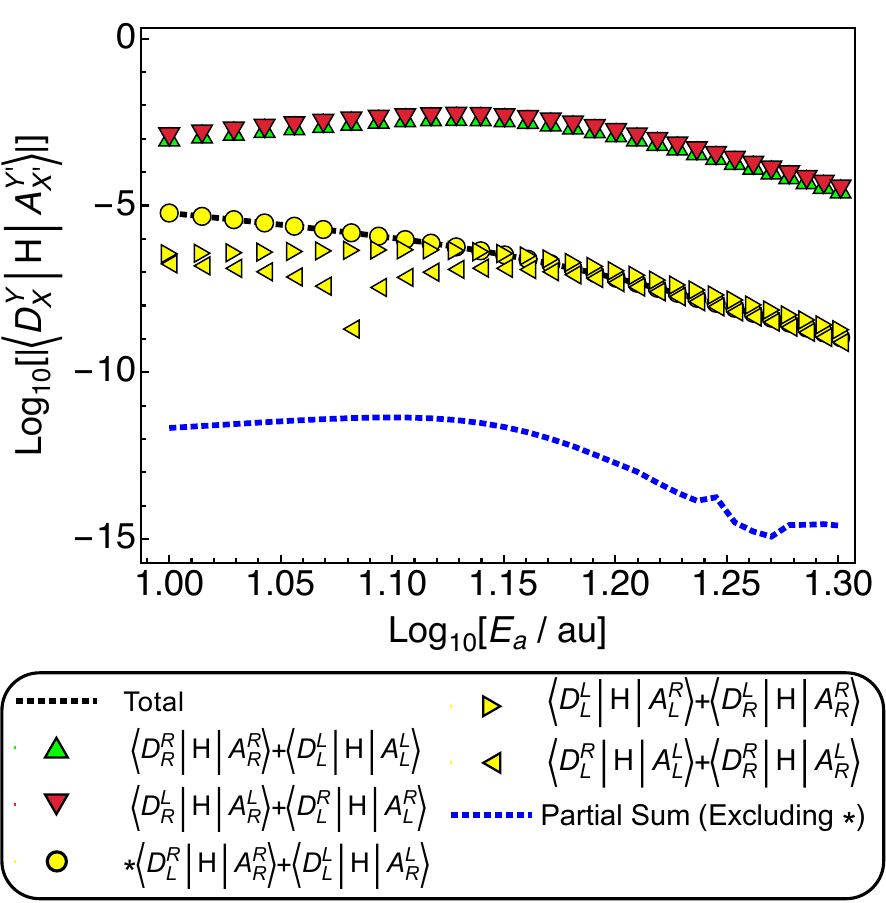}
    \caption{
    A log-log plot of the diabatic coupling decomposition based on excitation type as denoted by Eq.~\ref{eq:diabH} versus the barrier height. Note that the black dashed line is the total magnitude of the diabatic coupling, and summing all the terms except the term denoted with a preceding asterisk ($*$) results in negligible contribution to the sum. Thus, ($*$) is the only term that survives.
    }
    \label{fig:H_decompose}
\end{figure}

In Figure \ref{fig:H_decompose}, we show a log-log plot of the absolute value magnitude of the 16 terms in the sum for the diabatic coupling. 
Since  $V_{ext}$ is a symmetric double well potential,  the `matrix' in Figure \ref{fig:DXXHAXX_table} is persymmetric--meaning the term at (i,j) is equal to (5-j,5-i). Thus, there are only 10 unique terms to plot in Figure \ref{fig:H_decompose}. Note that the border color of the terms in the `matrix' of Figure \ref{fig:DXXHAXX_table} correspond to the color of the marker in Fig.~\ref{fig:H_decompose} and that individual terms with coupling less than $10^{-10}$ Hartree coupling are not included in the plot in Fig.~\ref{fig:H_decompose}.
At first glance, the largest individual terms 
$\bra{D_R^R} H \ket{A_R^R} + \bra{D_L^L} H \ket{A_L^L}$ (green triangle)
and 
$\bra{D_R^L} H \ket{A_R^L} + \bra{D_L^R} H \ket{A_L^R}$ (red triangle) in Fig.~\ref{fig:H_decompose}
appear significant. Interestingly, not only are these terms the largest terms by 2 orders of magnitude, but they are also much larger as compared to the total coupling shown by the black dashed line. That being said, these terms are are of opposite sign and largely cancel, and do not effectively contribute to the total coupling.

Apart from the red and green curves in Fig.~\ref{fig:H_decompose} (just discussed), 
the remaining three curves  in Fig.~\ref{fig:H_decompose} are plotted in yellow.  Here, the common color indicates that all of these curves result from a cross term between a local excitation on one diabat and a charge transfer on the other. Noticeably, 
$\bra{D_L^R} H \ket{A_R^R} + \bra{D_L^L} H \ket{A_L^R}$ (yellow circle)
is effectively on top of the curve for the total coupling, while 
$\bra{D_L^L} H \ket{A_L^R} + \bra{D_R^L} H \ket{A_R^R}$ (yellow right triangle)
and 
$\bra{D_L^R} H \ket{A_L^L} + \bra{D_R^R} H \ket{A_R^L}$ (yellow left triangle)
have $\sim 1$ order of magnitude smaller magnitude (and are of different signs and largely cancel).
In fact, 
the most striking feature of Fig.~\ref{fig:H_decompose} is that, if we sum up all of the the 16 terms in Eq.~\ref{eq:diabH} \emph{except}  the  $\bra{D_L^R} H \ket{A_R^R} + \bra{D_L^L} H \ket{A_L^R}$ (yellow circle), the resulting curve is effectively zero (plotted in blue).
Thus, we have arrived at two conclusions.
First, the tails of the diabatic wavefunctions are indeed the key and only important contributors to the overall diabatic coupling.
Second, not all of the possible tails for the diabatic state 
($\alpha, \beta, \gamma $and $\delta$  in Eq. \ref{eq:DTET_tails_guess}) are necessary. In fact, the diabatic coupling emerges from $\beta$ in this case, indicating that the diabatic coupling for a TET process emerges from the product of one charge transfer process (that establishes the size of $\beta$) and another charge transfer process that establishes the magnitude of the matrix element for tunneling from orbital $\bar{a}$ through orbital $\bar{d}$:

\begin{eqnarray}
    \left<D_{\mathrm{TET}} \middle| H \middle| A_{\mathrm{TET}}\right> \approx 
    \beta \left<
    \Phi_0'dba\bar{a}
    \middle | H | \Phi_0' d \bar{d} a b  \right> 
    \label{eq:final}
\end{eqnarray}

This result is of course the scaling law invoked by Closs {\em et al} though now made clear by calculation.
Indeed, if we set $\beta = H_{\textrm{DA}}^{\mathrm{ET}}/\Delta E$ by perturbation theory, then Eq.~\ref{eq:final} matches Eq.~\ref{eq:superexhange}. Note that this result would imply that, in order to model TET processes correctly and include electronic tails, it is essential to use modern, many-body electronic structure theory followed by an adiabatic to diabatic transformation. One might expect other methods based on constructing diabatic states directly (e.g. constrained DFT) to lack such tails and so  extracting robust TET diabatic couplings may be quite difficult (if not impossible)\cite{kaduk2012constrained, wu2006constrained, wu2006extracting, yeganeh2010triplet}. This hypothesis will need to be vigorously checked in the future.

\section{Discussion: Breakdown of Rate Relation in Closs Molecules}\label{sec:discussion}

\subsection{Molecules and Electronic Structure}\label{sec:discussion-calculations}

Let us now return to the Closs series of molecules that originally investigated TET and established the relationship between TET and ET/HT.
We can use the same techniques from above (for the model system) to analyze the Closs systems. Namely,
for ET calculations,  we performed  eDSC  simulations, described in Sec. \ref{sec:1D-hDSC} using a DIIS-SQP algorithm with the partitioning of the donor and acceptor fragments from Figure \ref{fig:molecule}. Similarly, we used hDSC to model HT.  All calculations were converged by requiring the norm of the gradient  to be less than $10^{-5}$au, and the temperature parameter for the dynamical weighting coefficients was set to 2eV.
All TET calculations were performed with CIS.  
 All calculations were run with a 6-31G* basis.

Recall that the original series of molecules came in two flavors, equatorial-equatorial attachment of the donor and acceptor to the cyclohexane (or dicyclohexane) bridge; or equatorial-axial attachment of the donor and acceptor to the bridge.  
As 
discussed  in the introduction of this work, experimentally it is well-established that the `ee' Closs series demonstrate a clear distance dependence with TET, HT, and ET rates scaling as $k=k_0e^{-\beta R_{\textrm{DA}}}$ as well as the aforementioned rate relation such that the relative rates for TET and ET rates both decay exponentially with $\beta_{\mathrm{TET}}=2\beta_{\mathrm{ET}}$.
We begin our analysis by computing the TET, ET, and HT diabatic couplings for the `ee' molecules measured experimentally. We find that these trends can be easily replicated using the theory described above: The slopes of the  ET/HT/(ET+HT) calculated plots are -1.18/-1.42/-2.60 (Fig. \ref{fig:closs_exp}d-f) vs. experiment (Fig. \ref{fig:closs_exp}a-c): -1.28/-1.50/-2.61.

Next, let us address  the `ae' and `ea' series of molecules. Here,  the
same robust measurement and analysis is not observed; in fact, for some of these molecules, the TET rates do not fall off exponentially with distance. Noticeably, also, the rates of C13ea an C13ae are not reported separately in Ref. \citenum{closs1988EET}. 
In a corresponding paper (Ref. \citenum{eshake}), we discuss these findings, which ultimately boil down to two factors (1) the breakdown of the Condon approximation by which the coupling fluctuates strongly with geometry, and for the C-13ae molecule, (2) there exists a thermodynamically accessible conical intersection (CI).

At this point, an intriguing question emerges: Given the hypothetical relationship established above, relating the TET rate to the product of the HT and ET rates, 
what are the implications of the presence of a CI for a TET process as far as predicting  ET and HT rates?
This question can now be answered  (at least in theory) by investigating  the C-13ae molecule. 
To that end, we selected 300 random points each for C-13ae and C-13ea from the data in Ref. \citenum{eshake} from E-SHAKE constrained sampling trajectories to explore the energetically accessible portions of the diabatic seam.
Additionally, to focus on the vanishing of the coupling in the presence of a CI, we select all geometries with less than \num{1e-8} au diabatic coupling as well ($\sim$260 geometries).
We evaluate the ET and HT diabatic coupling for these geometries as described above.

\begin{figure}[htp!]
    \centering
        \includegraphics[width=\linewidth]{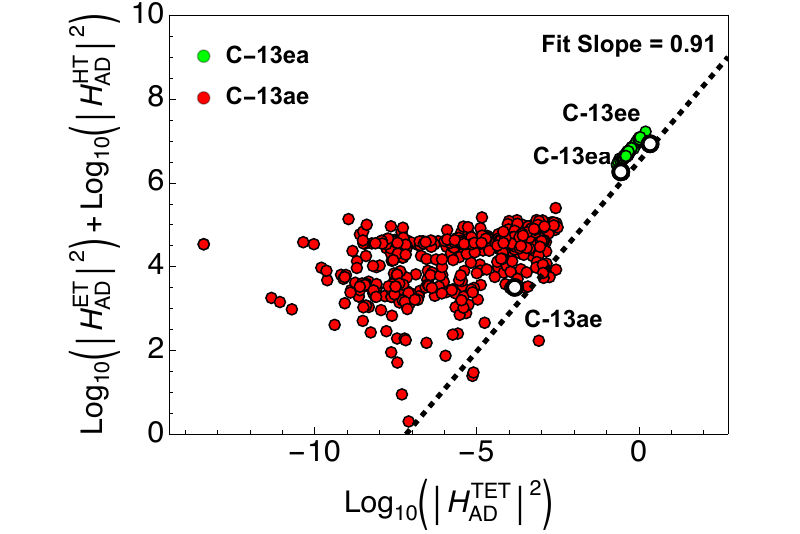}
    \caption{A scatter log-log plot of the squared diabatic coupling between D and A states for HT, ET (eDSC/hDSC), and energy transfer (CIS triplets) for the C-13ae (red) and C-13ea (green) molecules as shown in Figure \ref{fig:molecule}. The  open markers report the coupling at the optimized ground state geometry. The (black dashed) is a linear fit of the optimized ground state `ee' geometries.
    }
\label{fig:seam}
\end{figure}

In Figure \ref{fig:seam} we  compare the diabatic coupling for TET  to the HT/ET coupling (calculated using eDSC/hDSC) in a log-log plot for the C-13ae (red) and C-13ea (green) molecules. 
For comparison,  the black dashed line in Fig.~\ref{fig:seam} is a linear fit of the couplings calculated at the optimized ground state (singlet) `ee' geometries, with a slope of 0.91.
Deviations from this line illustrate a breakdown of the rate relationship given by Figure \ref{fig:et_ht_eet_drawing}.
While the data points sampled for the C-13ea molecule are in strong agreement with the simple rate relation exhibited by the `ee' molecules, the same is not the case for C-13ae.
To that point, take note of the farthest deviating point for the C-13ae which is incredibly far from  from the linear trend (approx. 13 orders of magnitude). Using a Marcus treatment of the rate this data point would suggest energy transfer is nearly zero to machine precision while the electron and hole transfer couplings are finite and are from a far narrower distribution.
It follows that that the TET vs HT/ET scaling law fails dramatically when in the presence of a CI. Moreover, given the sensitivity to the TET diabatic coupling expression to perturbation theory (and the need for two small interactions---one generating a wavefunction tail and the other an interaction of that tail itself), it stands to reason that small TET diabatic couplings will fluctuate far more than ET diabatic couplings (which are dominated by only single small interaction), so that the universal nature of the such a scaling law may be limited for less rigid molecules.

\section{Conclusion}

In summary, we have built a simple 1D model for ET and TET dynamics to model ET and TET rates, and using this model we have shown how the scaling law ($k = k_0 e^{-\beta R_{\textrm{DA}} }$ with $\beta_{\mathrm{TET}}  = 2\beta_{\mathrm{ET}}$) emerges from  the existence of weak tails in the TET diabatic states formed from an ATD transformation.  The sensitivity of the TET diabatic coupling to small wavefunction tails  would suggest that TET processes may well be sensitive to large fluctuations in the diabatic coupling (moreso than ET or HT processes). Moreover, using the `ae' and `ea' Closs molecules (where the TET coupling is known to be highly sensitive to nuclear geometry), we showed that the TET vs HT/ET relationship will break down whenever the TET couplings fluctuates strongly (as the ET/HT couplings do not appear as sensitive).

Looking forward, our predictions above can be quantified within a standard linear coupling model\cite{zimmermann1985bilinear}, for which a typical two-state nonadiabatic model for an electron or energy  transfer process can be written in the form:
\begin{eqnarray}
    H = \frac{1}{2}k_X \bX^2 + 
    \frac{1}{2}k_Y \bY^2 +  \left(
    \begin{array}{cc}
    \epsilon_a  + \lambda \bX & V_0 + \eta \bY \\
    V_0 + \eta \bY &  \epsilon_b -\lambda \bX
    \end{array}
    \right)
\end{eqnarray}
The statement above that the TET diabatic coupling should be more geometry dependent than the ET diabatic coupling would suggest that $|\eta_{\mathrm{TET}}| > |\eta_{\mathrm{ET}}|$.
That being said, however, because charge transfer involves a bigger net change in electric field to the environment, one would expect that $|\lambda_{\mathrm{ET}}| > |\lambda_{\mathrm{TET}}|$. 
Such a hypothesis can certainly be tested with modern electronic structure theory that fits such linear couplings models routinely, and our hope is that the results will yield new insights into the still not fully understood nature of nonadiabatic processes.

\begin{acknowledgement}

This work was supported by the U.S. Air Force Office of Scientific Research (USAFOSR) (under Grant No. FA9550-23-1-0368). 
We thank the DoD High Performance Computing Modernization Program for computer time.
This material is based upon work supported by the National Science Foundation Graduate Research Fellowship under Grant No. DGE-2236662.
\end{acknowledgement}



\bibliography{main}

\end{document}